\documentclass[aip,reprint]{revtex4-1}

\usepackage{graphicx}
\usepackage{dcolumn}
\usepackage{bm}
\usepackage{braket}
\usepackage{appendix}
\usepackage{todonotes}
\usepackage{soul}
\usepackage{changes} 
\usepackage[colorlinks = true, linkcolor = blue, citecolor = blue, urlcolor = blue]{hyperref}
\usepackage{academicons}
\usepackage{minted}
\usepackage{comment}
\usepackage{subfiles}
\usepackage{changepage}
\usepackage{siunitx}
\DeclareSIUnit{\dBm}{\deci\belmilliwatt}
\DeclareSIUnit{\dB}{\decibel}
\usepackage{caption}
\usepackage{subcaption} 
\usepackage{amsmath, amssymb}

\AtBeginDocument{
  \heavyrulewidth=.08em
  \lightrulewidth=.05em
  \cmidrulewidth=.03em
  \belowrulesep=.65ex
  \belowbottomsep=0pt
  \aboverulesep=.4ex
  \abovetopsep=0pt
  \cmidrulesep=\doublerulesep
  \cmidrulekern=.5em
  \defaultaddspace=.5em
}
\usepackage{booktabs, tabularx}
\usepackage{footmisc}

\usepackage{ragged2e} 
\newminted{python}{%
}

\newcommand{\orcid}[1]{\href{https://orcid.org/#1}{\textcolor[HTML]{A6CE39}{\aiOrcid}}}

\draft 

\begin{document}

\preprint{APS/123-QED}


\title{Operation of Unshielded Kinetic-Inductance Traveling-Wave Parametric Amplifiers in Multi-Tesla Fields} 




\author{C. Boutan}
\affiliation{Pacific Northwest National Laboratory, 
            902 Battelle Blvd., 
            Richland, WA 99352, USA}

\author{E. Lentz}
\affiliation{Pacific Northwest National Laboratory, 
            902 Battelle Blvd., 
            Richland, WA 99352, USA}

\author{C. Shiu}
\affiliation{Quantum Sensors Division, National Institute of Standards and Technology, Boulder, Colorado 80305, USA}

\author{D. Erdag}
\affiliation{Pacific Northwest National Laboratory, 
            902 Battelle Blvd., 
            Richland, WA 99352, USA}

\author{S. Jones}
\affiliation{Pacific Northwest National Laboratory, 
            902 Battelle Blvd., 
            Richland, WA 99352, USA}

\author{J. Van Vlack}
\affiliation{Pacific Northwest National Laboratory, 
            902 Battelle Blvd., 
            Richland, WA 99352, USA}

\author{L. Howe}
\affiliation{California Institute of Technology, Pasadena, California 91125, USA}
\affiliation{Quantum Sensors Division, National Institute of Standards and Technology, Boulder, Colorado 80305, USA}
\affiliation{Department of Physics, University of Colorado, Boulder, 80309, Colorado, USA}

\author{A.~Giachero}
\affiliation{Quantum Sensors Division, National Institute of Standards and Technology, Boulder, Colorado 80305, USA}
\affiliation{Department of Physics, University of Colorado, Boulder, 80309, Colorado, USA}
\affiliation{Department of Physics, University of Milano Bicocca, Milan, I-20126, Italy}

\author{P. Szypryt}
\affiliation{Quantum Sensors Division, National Institute of Standards and Technology, Boulder, Colorado 80305, USA}
\affiliation{Department of Physics, University of Colorado, Boulder, 80309, Colorado, USA}

\author{M. Vissers}
\affiliation{Quantum Sensors Division, National Institute of Standards and Technology, Boulder, Colorado 80305, USA}

\author{J. Austermann}
\affiliation{Quantum Sensors Division, National Institute of Standards and Technology, Boulder, Colorado 80305, USA}

\author{J. Hubmayr}
\affiliation{Quantum Sensors Division, National Institute of Standards and Technology, Boulder, Colorado 80305, USA}

\author{S. Knirck}
\affiliation{Department of Physics, Harvard University, Cambridge, Massachusetts 02138, USA}

\author{D. A. Bennett}
\affiliation{Quantum Sensors Division, National Institute of Standards and Technology, Boulder, Colorado 80305, USA}
\affiliation{Department of Physics, University of Colorado, Boulder, 80309, Colorado, USA}

\author{J. Ullom}
\affiliation{Quantum Sensors Division, National Institute of Standards and Technology, Boulder, Colorado 80305, USA}
\affiliation{Department of Physics, University of Colorado, Boulder, 80309, Colorado, USA}



\date{\today}

\begin{abstract}

Cryogenic parametric amplifiers are used to amplify radio-frequency signals for a range of applications in basic and applied science. 
Both Josephson Parametric Amplifiers and Josephson Traveling-Wave Parametric Amplifiers have been used as first-stage amplifiers enabling readout chains operating within a few quanta of the quantum limit. However, these devices are highly sensitive to magnetic fields, having critical current suppressed by the Fraunhofer effect, requiring substantial field-free zones. 
In a dark matter axion search experiment, axions convert to detectable microwave photons in the presence of a strong magnetic field, necessitating amplifiers that can reliably operate close to these environments. 
Kinetic-inductance Traveling-Wave Parametric Amplifiers (KTWPAs) may be the ideal candidate for this type of application having high critical magnetic field of the materials used throughout their construction. In this letter we demonstrate that KTWPAs can provide high gain ($>$20~dB) over a multi-GHz bandwidth in spite of from multiple exposures to multi-Tesla fields. Further, we explore operational characteristics of these devices under harsh conditions as a function of overall field strength, device orientation within the field, applied bias current, and pump power \& frequency. In so doing, we find KTWPA gain vanishes in devices oriented perpendicularly to a field of 0.02~\si{T}, but gain values $>$10~dB are achievable in fields over 1~\si{T} when oriented near~parallel to the device plane, with peak gain achieved with an applied 0.25~\si{T} to 0.5~\si{T} field. It is our expectation that KTWPAs will expand the accessibility of quantum-limited RF measurements in the presence of Tesla-scale fields.

\end{abstract}


\maketitle 


Radio-frequency (RF) amplifiers capable of significantly boosting signals with introduced noise at or near the standard quantum limit (SQL) are becoming increasingly sought after for basic science~\cite{ASZTALOS201139,Du_2018,PhysRevLett.124.101303,Tatsumi_2021,ahn2024extensivesearchaxiondark,Youn_2024} and even more heavily for applications including quantum computing~\cite{abdo2011}. 

Today's most widely used quantum-limited pre-amplifiers are the Josephson Parametric Amplifiers (JPAs). Utilizing the non-linear inductance of Josephson Junctions (JJs), these amplifiers have exquisite sensitivity with added noise down to the minimum allowed by quantum mechanics \cite{yurke1989observation,castellanos2007widely}. The basic architecture, a nonlinear JJ element embedded in a superconducting resonator, however has inherent limitations which present challenges to broad application. 

First, the instantaneous bandwidth of the amplifier is set by the resonance bandwidth and is typically a few tens of megahertz. While tunable across gigahertz scales, moving to a different operation frequency requires extensive retuning. Impedance matching~\cite{mutus2014strong} and filter synthesis~\cite{kaufman2023josephson} techniques have demonstrated significant increases in bandwidth but become increasingly challenging at higher frequencies~\cite{Hao_2026}. 
Josephson junction critical currents are suppressed by the Fraunhofer effect, making their operation sensitive to $\mathcal{O}(\unit{mT})$ fields \cite{Janssen2024}. 
Furthermore, the same critical current also sets the tuning scale, limiting dynamic range~\cite{Castellanos_Beltran_2009}.

The bandwidth limitation has been addressed by moving from a resonator to transmission line geometry. This creates a Traveling-Wave Parametric Amplifier (TWPA) with an instantaneous bandwidth set not by the resonator, but by the extent of phase matching between pump, signal, and idler tones. Still using Josephson junctions, TWPAs can be fabricated with nearly an octave of instantaneous bandwidth and added noise down to the SQL~\cite{kevinobrien}.

Kinetic inductance in a superconductor arises from the inertia of Cooper pairs, and is naturally a non-linear element. The magnetic field resilience of kinetic inductance devices over junction-based devices results from the higher critical fields of type II superconductors. The kinetic inductance per unit length ($\mathcal{L}_K$) of a superconducting film as a function of current, $I$, is nontrivial~\cite{Kubo2020} but can be Taylor-expanded in the small-signal limit in even powers of the net device current $I$ \cite{Anlage1989} as
\begin{equation}
    \mathcal{L}_K(I) = \mathcal{L}_0 \left[ 1 + \left(\frac{I}{I_{*,2}}\right)^2 + ... \right],
\end{equation}
where $\mathcal{L}_0$ is the zero-bias and zero-frequency kinetic inductance per unit length and $I_{*,2}$ is the second order scaling current. $I_{*,2}$ parametrizes the scale of the non-linearity, and is a function of superconducting material and geometry \cite{Giachero_2023}. 

In this letter, we demonstrate that KTWPAs can maintain high gain in the presence of substantial in-plane magnetic fields. Magnetic field resilience has not previously been studied with NIST-fabricated devices. 
Janssen \textit{et. al.} demonstrated similar measurements with devices fabricated at JPL \cite{Janssen2025}, which use similar materials. Here, we reproduce and extend these results with an independent fabrication and experimental platform. Our results show higher gain over a band extending to higher frequencies, above 9~GHz, and while maintaining gain at higher magnetic field strengths when oriented parallel to the device plane. 

\begin{figure}[h!]
    \centering
    \includegraphics[width=0.5\textwidth]{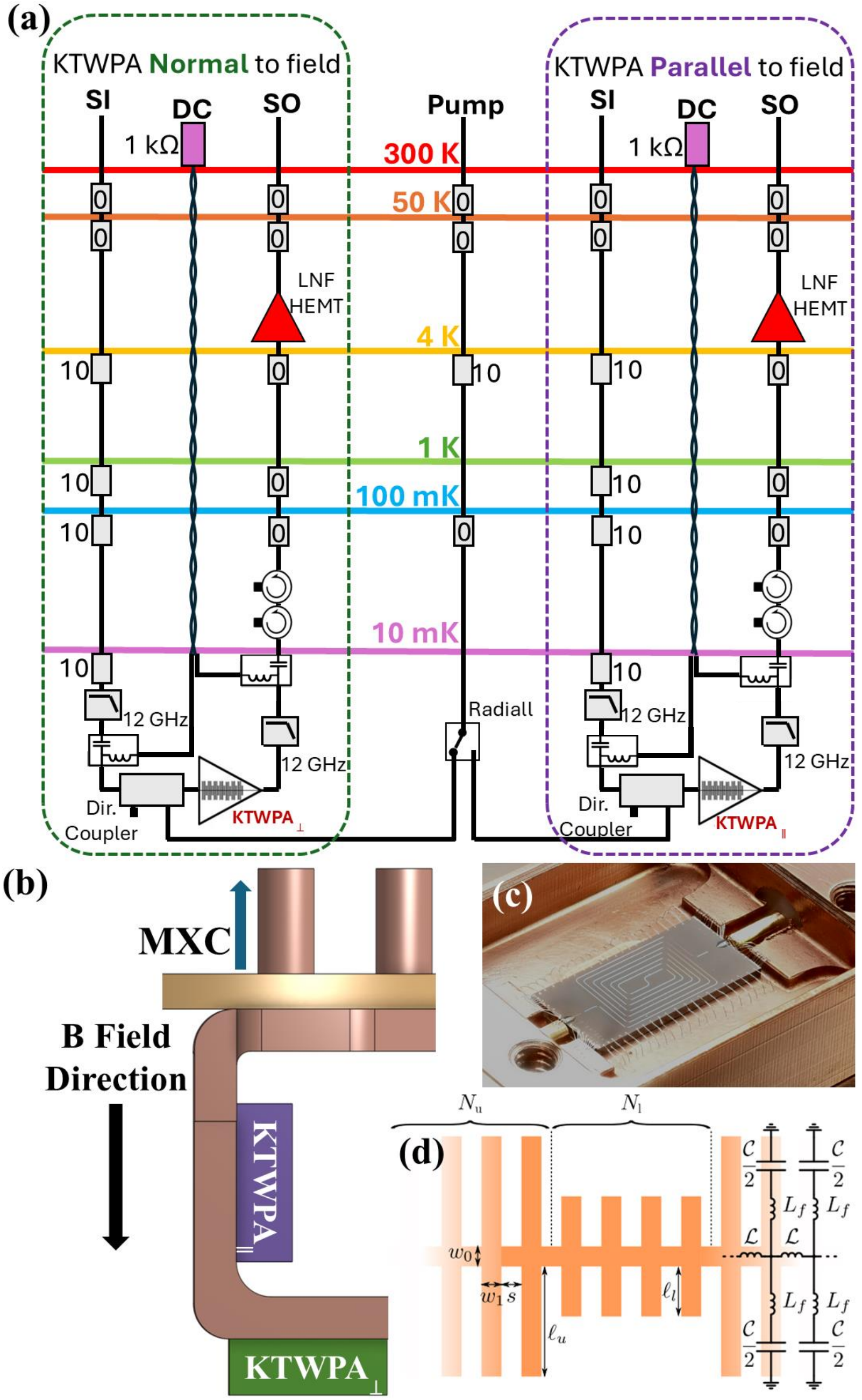}
    \captionsetup{format=plain,justification=centering} 
    \caption{\justifying(A) Schematic of DC and RF wiring for experiment (without the magnet). (B) Schematic showing the KTWPA orientation relative to the applied magnetic field. (C) Image of packaged device, with chip dimensions 5$\times$10 $\si{mm}^2$. (D) KTWPA supercell schematic, reproduced with permission from~\cite{howe2025kineticinductancetravelingwave}. }
    \label{fig:schem}
\end{figure}

This letter exercises two KTWPAs similar to those described in Howe \textit{et al.}~\cite{malnou_2021, howe2025kineticinductancetravelingwave}, to which we refer the interested reader for additional details on the design, modeling, and general characterization. As shown in \figureautorefname{ \ref{fig:schem}c \& d},  
the KTWPA devices use a stub-loaded inverted microstrip configuration.  NbTiN is the central conductor film with thickness 10~\unit{\nano \meter}  ($\mathcal{L}_k$ = 30pH/$\square$) and width $w_0 = 1$~\unit{\micro \meter} deposited on a Si substrate.
It is then coated in 100~\unit{\nano \meter} amorphous-Si and 200~\unit{\nano \meter} Nb films, forming a continuous ground plane. ``Quarter-wave'' stubs are added to engineer an artificial transmission line, creating a dispersion relationship favorable for three-wave mixing gain processes. 

\par
Our experimental schematic is shown in \figureautorefname{ \ref{fig:schem}a} and further details can be found in Supplementary Information. The amplifier is configured to support three-wave mixing, with bias-tees that bring in a DC bias and a combination of directional couplers and filters to add and remove pump photons. The amplifiers are mounted on a copper bracket whose orientation is parallel and perpendicular to the AMI 9+1+1 Tesla vector magnet. The alignment of the device was established by mechanical assembly. No magnetic orientation sensors were installed in-situ.


Critical current is estimated by discontinuities in the DC IV response, and summarized in Table~\ref{tab:currents}. These measurements establish the accessible operating bias of the device under a range of magnetic field strengths. A weak enhancement of the critical current is observed at low fields before monotonically decreasing at higher fields.

\begin{table}
\caption{Critical current ($I_c$) measurements under different magnetic fields for both devices.}
\label{tab:currents}
\centering
\begin{tabular}{cccc} 
\toprule
\multicolumn{2}{c}{KTWPA$_\perp$} & \multicolumn{2}{c}{KTWPA$_\parallel$} \\
\cmidrule(lr){1-2}
\cmidrule(lr){3-4}
Field (mT) & $I_c$ (\si{\micro\ampere}) & Field (T) & $I_c$ (\si{\micro\ampere})  \\ 
\midrule
0 & 400 $\pm$ 5 & 0 &  490 $\pm$ 5 \\ 
 5 & 460 $\pm$ 5 & 0.25 & 530$\pm$ 5 \\ 
 10 & 300 $\pm$ 5 & 0.50 & 400 $\pm$ 5 \\ 
 15 & 330 $\pm$ 5 & 0.75 & 540 $\pm$ 5 \\
 20 & 260 $\pm$ 5 & 1.0 & 450 $\pm$ 5 \\ 
 25 & 250 $\pm$ 5 & 1.25 & 440 $\pm$ 5 \\ 
 30 & 240 $\pm$ 5 & & \\ 
 35 & 230 $\pm$ 5 & & \\ \bottomrule
\end{tabular}

\end{table}

\begin{figure*}[ht!]
\centering
\begin{subfigure}{0.49\textwidth} 
\includegraphics[width = \textwidth]{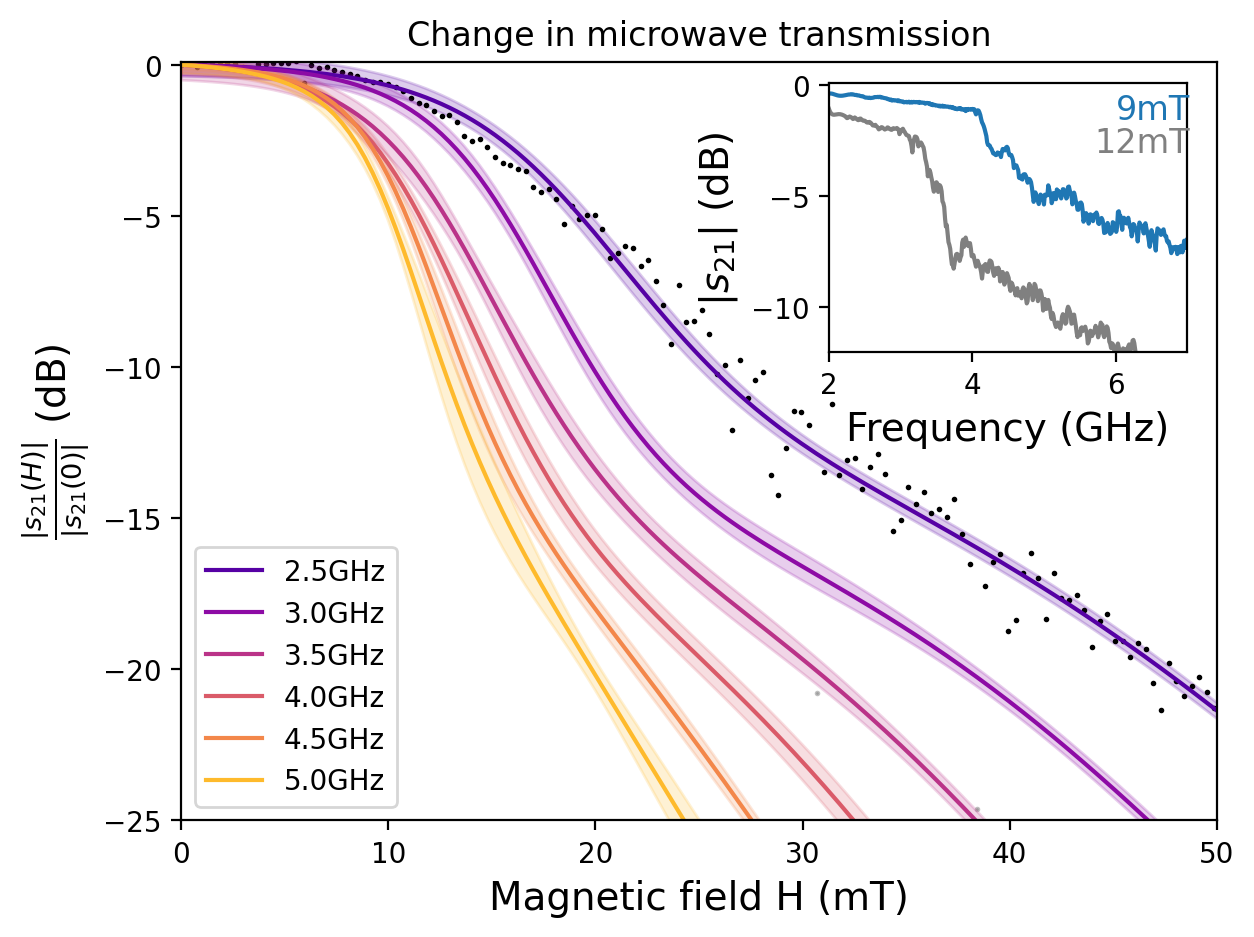}
\end{subfigure}
\begin{subfigure}{0.49\textwidth} 
\includegraphics[width = \textwidth]{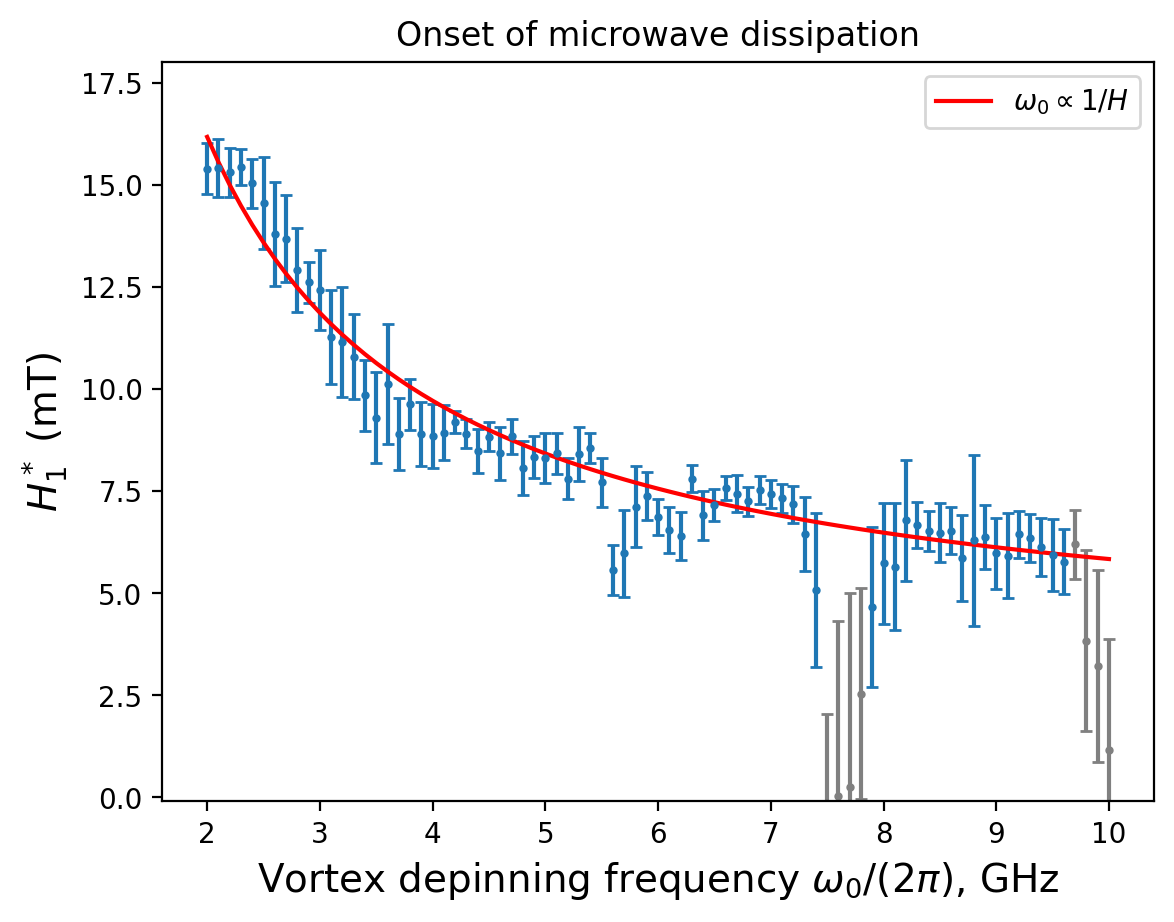}
\end{subfigure}
\caption{(Left) Change in microwave transmission of KTWPA$_\perp$ during a field ramp-down. For clarity, only one set of data is shown (gray dots, $2.5\,\si{GHz}$), with model bands shown by the color fits. A two-component sigmoid model is used to represent the data. (Inset) microwave transmission as a function of frequency, for fixed magnetic field strength ($9,12\,\si{mT}$). A distinct inflection in microwave loss is visible, consistent with the GR model for vortex depinning. However, it appears that the loss mechanism softens above this inflection. 
(Right) Onset of vortex depinning as a function of magnetic field strength and microwave frequency. For each frequency, the magnetic field inflection point, $H_1 - \sigma_1$ is extracted by the fit, as shown in the left panel. A theoretical $1/H$ curve dependence is included as a guide to the eye, which captures the qualitative trend of the data. The gray points indicate unreliable inferences due to package modes.}
\label{fig:magfieldss21}
\end{figure*}

The KTWPA is constructed from two different superconducting materials, NbTiN with a Nb ground plane, and these materials are expected to exhibit different magnetic responses. 
NbTiN is a highly disordered superconductor with a large Ginzburg-Landau (GL) parameter $\kappa = \lambda_L/\xi_{GL} > 50$, where it has been reported that NbTiN has a London penetration depth of $\lambda_L^{\text{NbTiN}} = 200\,\si{nm}$ to $380\,\si{nm}$\cite{Yu_2005, Khan_2023, Lee_2024} and a $\xi_{GL} = 3.8\,\si{nm}$ \cite{Yu_2002, Lee_2024}. However, exact growth conditions and stoichiometry can cause these values to vary. 
Nevertheless, the large penetration depth implies a lower critical field $H_{C1} \approx 18\,\si{mT}$, allowing magnetic vortices to enter the film. 
This material also has a short coherence length, leading to superconductivity persisting to higher fields $H_{C2}$, estimated to be $\sim 22\si{T}$ \cite{Lee_2024, Muller_2022}.

In contrast, bulk Nb lies near the type-I/type-II boundary ($\lambda_L = 29\,\si{nm}, \xi_{GL} = 40\,\si{nm}, \kappa \approx 1/\sqrt{2}$ \cite{Maxfield_1965}). In sputtered Nb films, disorder increases $\lambda_L \approx 70\,\si{nm}$ to $90\,\si{nm}$, resulting in type-II behavior \cite{Gubin_2005}, and a small vortex mixed state. Reports have indicated $H_{C1} \approx  170\,\si{mT}$. \cite{Maxfield_1965, Finnemore_1966, McFadden_2026}.

In our devices, the fields penetrate the superconducting film, changing the complex-conductivity and ultimately limiting device performance. We first focus on the real component, where the magnetic field creates magnetic vortices that cause dissipation. 
At a fixed microwave frequency, change in transmission vs. applied out-of-plane magnetic field for KTWPA$_{\perp}$ is shown in Fig. \ref{fig:magfieldss21} (left). 
Two distinct transitions are visible in the microwave transmission, suggesting two vortex populations. This is consistent with a two-film device with different energy regimes. Therefore, we fit our model to a heuristic model, a sum of two sigmoids, 
\begin{equation}
|S_{21}| (H) = \frac{A}{1 + \exp \left( \frac{H - H_1}{\sigma_{H_1}}  \right)} + \frac{B}{1 + \exp \left(\frac{H - H_2}{\sigma_{H_2}} \right) } + C.
\label{eq:s21_bfield_fit}
\end{equation}
$H_1, H_2$ represent two transitions in microwave transmission along with the respective widths, $\sigma_{H_1},\sigma_{H_2}$. These are unrelated to the thermodynamic critical fields $H_{C1}, H_{C2}$.

Evident in Fig. \ref{fig:magfieldss21}(left, inset), the microwave transmission diminishes with a sharp cutoff. 
This phenomenon can be described by the Gittleman and Rosenblum (GR) model \cite{Gittleman_1964, Calatroni_2019}, where vortices respond to microwave power to cause dissipation in different frequency regimes. This model treats the vortex as a damped harmonic oscillator, with a characteristic depinning frequency $\omega_0$. At microwave frequencies below the depinning frequency ($\omega < \omega_0$), pinning forces dominate and vortex motion is nearly nondissipative. Microwave drives above $\omega > \omega_0$ depins the vortex, creating flux flow, and consequentially a steep increase in dissipation. 
Within the GR model, the depinning frequency is the ratio of pinning stiffness to flux viscosity $(\omega_0 = k/\eta)$, both of which can have field dependence. 
The observed loss, while near-monotonic, has a softening of dissipation at higher frequencies. This behavior is not captured by a simple GR description with a single $\omega_0$. One possible explanation is a saturation of the vortex flux-flow loss; however, a detailed mechanism is beyond the scope of this work.

In Fig. \ref{fig:magfieldss21} (right), the field-dependent transmission is fit to Eq. \ref{eq:s21_bfield_fit}, where we interpret $H_1^* \equiv H_1 - \sigma_{H_1}$ as the onset of vortex depinning. 
We compare the shape of the curve to $\omega_0 \propto H^{-1}$, and show a qualitative agreement with a decreasing $\omega_0$ with increasing applied field, although alternative powers may better capture the curvature of the data.
Some experiments have observed the field dependence of the depinning frequency \cite{Bonura_2008, Gittleman_1968}, while others have not exhibited a field-dependence \cite{Song_2009}. 
Plausible physical origins of the observed field dependence include a reduction of the pinning stiffness $k$ with increasing $H$-field. In collective pinning regimes with many vortices per pinning site, the vortices will have a field dependent pinning stiffness, and power-law dependencies $(k \propto H^{-\alpha}$ with $\alpha \in [0.5-1]$) have been discussed in the literature \cite{Golosovsky_1996, Gittleman_1968}.

\begin{figure*}[ht!]
\centering
\includegraphics[width = 0.5\textwidth]{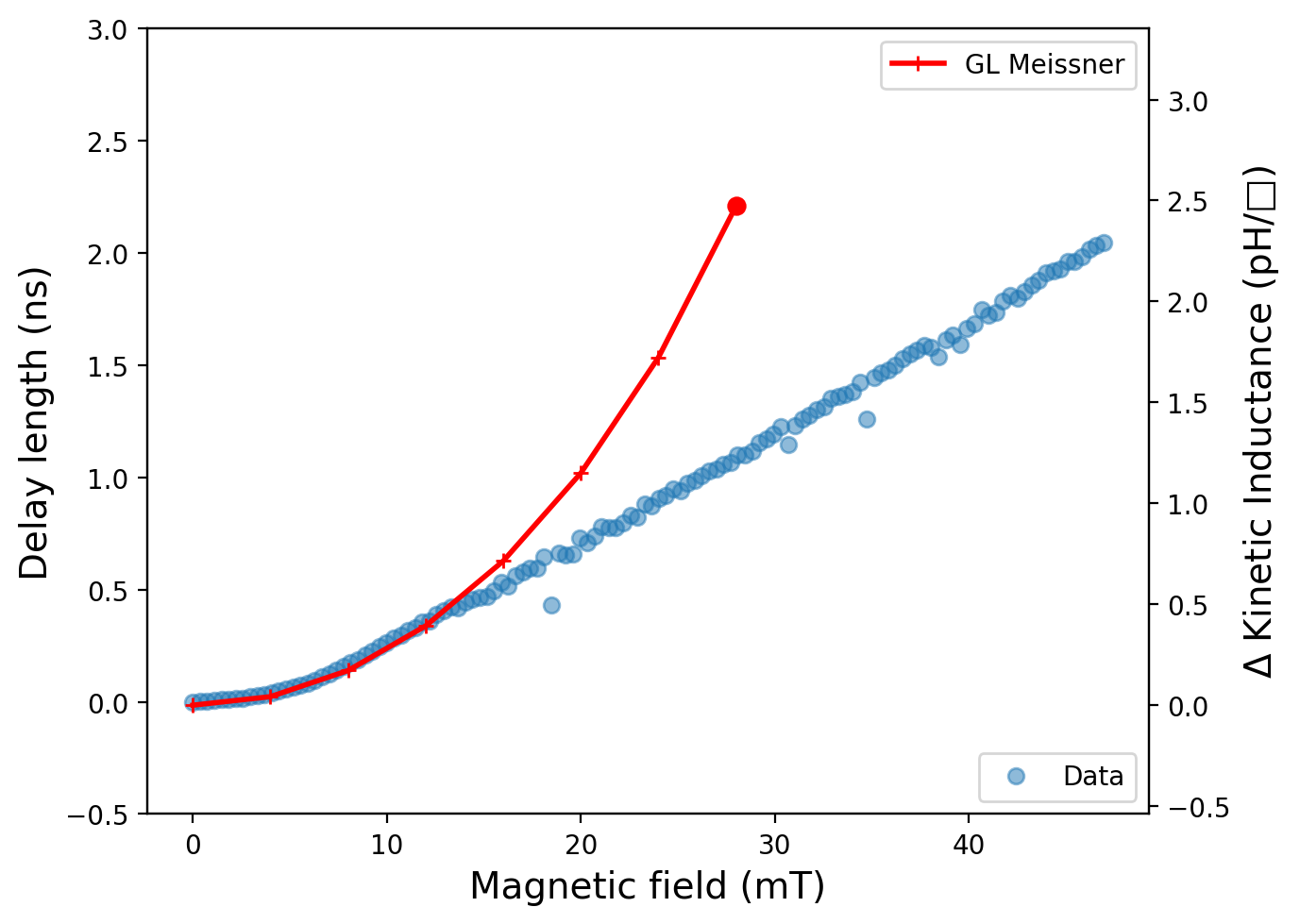}
\caption{Time-delay (left axis) of the KTWPA$_\perp$ as a function of magnetic field, converted to a change in kinetic inductance (right axis). A GL calculation of the kinetic inductance is shown in red, with $\xi = 3.9\,\si{nm}, \lambda_L = 450\,\si{nm}, $ and $\gamma = 10.$ The range of the simulation is restricted to entirely be in the Meissner state. Magnetic fields suppress the superconducting state, increasing the kinetic inductance, but difference in the shape of theory and experiment suggests additional mechanisms at play. Note the divergence in behavior at approximately $H_{C1}$ of 18~mT for NbTiN.} 
\label{fig:s21_phase}
\end{figure*}

The reactive term of the complex conductivity is encoded in the phase of $S_{21}$. In Fig. \ref{fig:s21_phase}, we measure the delay length of the KTWPA as a function of the out-of-plane magnetic field strength, normalized the length at zero field ($B_0 = 0$). We observe a steady increase in the electrical length of the device with increasing magnetic field. Assuming that the change in delay length is purely reactive, we infer the corresponding effective kinetic inductance using \texttt{TWPASolver}, a coupled mode equation solver \cite{TWPASolver}.

We can compare the change in kinetic inductance to numerical solutions of the Ginzburg-Landau (GL) equation under different magnetic field conditions using \texttt{pyTDGL} \cite{Bishop-Van_Horn_2023}. The total GL free energy $F(I, B_0)$ is calculated and the kinetic inductance is, 
\begin{equation} 
\begin{aligned}
&L_k (I_0, B_0) = \left. \frac{d^2 F(I, B_0)}{d I^2} \right|_{I = I_0} \\ 
&= \frac{F(I_0 + \Delta I, B_0) - 2F(I_0, B_0) + F(I_0 - \Delta I, B_0)}{\Delta I^2} \\
&+ O(\Delta I^3).
\end{aligned}
\end{equation} 
For simplicity we only consider the Meissner state shown by the red curve in Fig. \ref{fig:s21_phase}. Physically, screening currents suppresses superconductivity and increases the kinetic inductance. Both theory and experiment show this monotonic trend. However, GL predicts a nonlinear trend with applied magnetic field, whereas the data show a linear trend. This suggests that the observed increase does not arise purely from screening currents, but likely involves an additional mechanism. 
Other publications have considered the kinetic inductance of the vortex contribution which they found to contribute inductance on the order of $\sim 0.01$pH \cite{kalashnikov_2026}. 
Nevertheless, these calculations and data give a bound on the change in kinetic inductance of no more than 2.5pH/$\Box$ or a change less than 8\%. These calculations suggest that magnetic field penetration does not destroy the superconducting state, rather it perturbs the kinetic inductance to a higher value, which can be compensated for in the operation of the device.

Gain measurements are performed with an on/off measurement of the pump. 
Measurements were preceded by other measurements in which a large B-field was applied, and no attempt was made to remove any trapped flux that may have been created by those earlier measurements, detailed in Table \ref{table:magramps}.
A summary of magnetic fields, bias currents, pump frequencies, and pump powers explored during gain measurements for each KTWPA is shown in Table \ref{table:params}. The magnetic field direction is held constant throughout the data acquisition.

\begin{table}
\centering
\caption{Parameter space explored for KTWPA$_{\perp}$ and KTWPA$_{\parallel}$.}
\begin{tabular}{lll}
\toprule
Parameter & KTWPA$_{\perp}$ & KTWPA$_{\parallel}$ \\
\midrule
Magnetic field &
0--35 mT  &
0--1.5 T  \\
Bias current &
120--320 $\mu$A &
260--350 $\mu$A \\
Pump frequency &
12.5--13.5 GHz &
13.9--14.6 GHz \\
Pump power &
21--23 dBm &
21--23 dBm \\
\bottomrule
\end{tabular}
\label{table:params}
\end{table}

\begin{figure*}[hbt!]
    \centering
    \includegraphics[width=\textwidth]{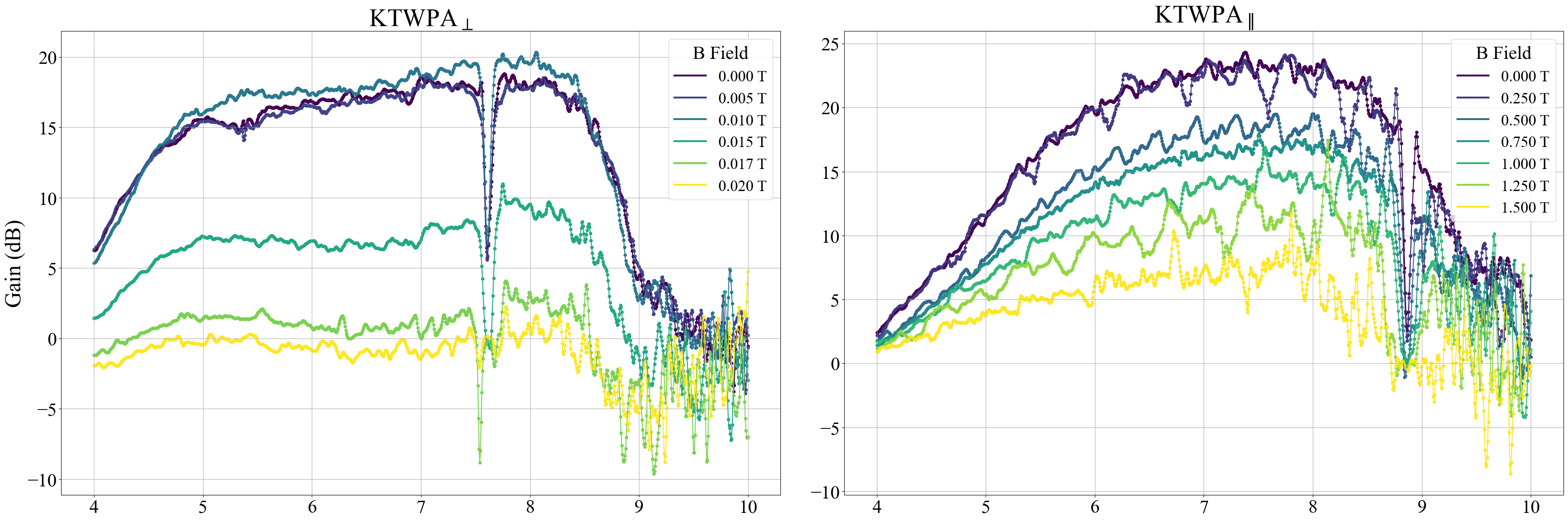}
    \includegraphics[width=\textwidth]{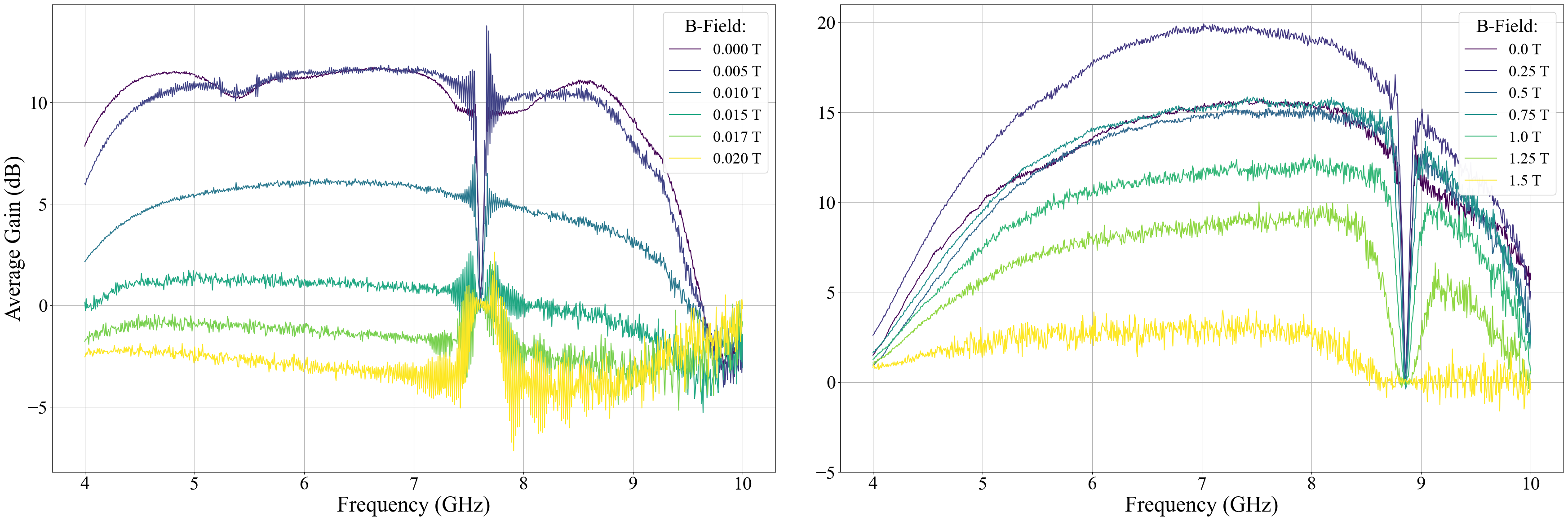}
    \captionsetup{format=plain,justification=centering} 
    \caption{\justifying \textit{Top row:} Gain recorded for both devices as a function of magnetic field for a set of fixed operating parameters. KTWPA$_\perp$ gain was collected with a pump frequency of \SI{13}{\giga\hertz}, bias current of \SI{220}{\micro\ampere}, and pump power of 22~dBm, and KTWPA$_\parallel$ with pump frequency \SI{14.2}{\giga\hertz}, bias current \SI{300}{\micro\ampere}, and pump power 22~dBm. The narrow ($\sim 100$~MHz) dip in gain at 7.7~GHz for KTWPA$_\perp$ and 8.9~GHz for KTWPA$_\parallel$ are substrate modes and not fundamental to the device's magnetic response.
    \textit{Bottom row:} We show the  gain attained by these devices at each tested external field, averaged over the pump frequency, pump power, and bias current of the device at each RF signal frequency tested. }
    \label{fig:b-field-lines}
\end{figure*}

Example on/off gain profiles for each KTWPA over their respective magnetic field values and fixed pump frequency, pump power and bias current are given in Fig.~\ref{fig:b-field-lines}. As expected, the magnetic field diminishes the overall gain performance of the device. However, their gain profiles show they remain proficient broadband amplifiers highly resilient to large degrees of flux (KTWPA$_\perp$) and high intensity fields (KTWPA$_\parallel$). 

The gain evolution with magnetic field cannot be explained solely by a dissipative mechanism; a pure dissipation would be invisible in an on/off measurement of gain. Additionally, as evidenced by Fig.~\ref{fig:b-field-lines}, the field does not substantially change the overall shape of the gain-bandwidth profile. This suggests that the magnetic field is not primarily acting through a uniform suppression of the kinetic inductance, which would cause a shift in the dispersion relation, the pump phase-matching condition, and ultimately change the shape of the gain profile. 
Our calculations reveal that kinetic inductance should change by no more than 8\%, which is within the range of operability of our devices.

The KTWPA$_\perp$ profiles overall show a resilience to about 5~\si{mT} to 10~\si{mT}, after which gain drops off significantly. By 20~mT nearly all the explored parameter settings produce no net gain. This is consistent with our bias-less microwave transmission measurements showing no transmission by 20~mT. This suggests that the vortex loss exceeds the parametric amplification process with this applied field. 
Compared to Janssen et. al \cite{Janssen2025}, our amplifiers show higher gain-suppression in out-of-plane magnetic fields, while preserving gain at higher in-plane magnetic field. 
The out-of-plane behavior may be attributed to our larger feature sizes ($1\si{\micro m}$ compared to their $0.34\si{\micro m}$\cite{faramarzi_2024}), as larger linewidths reduce the energy required for vortex entry. The in-plane behavior may be attributed to these amplifiers having thinner films ($10\si{nm}$ vs $35\si{nm} \cite{faramarzi_2024}$), which naturally enhances the upper critical field \cite{tinkham}.

\begin{figure*}
    \begin{center}
    \includegraphics[width=\textwidth]{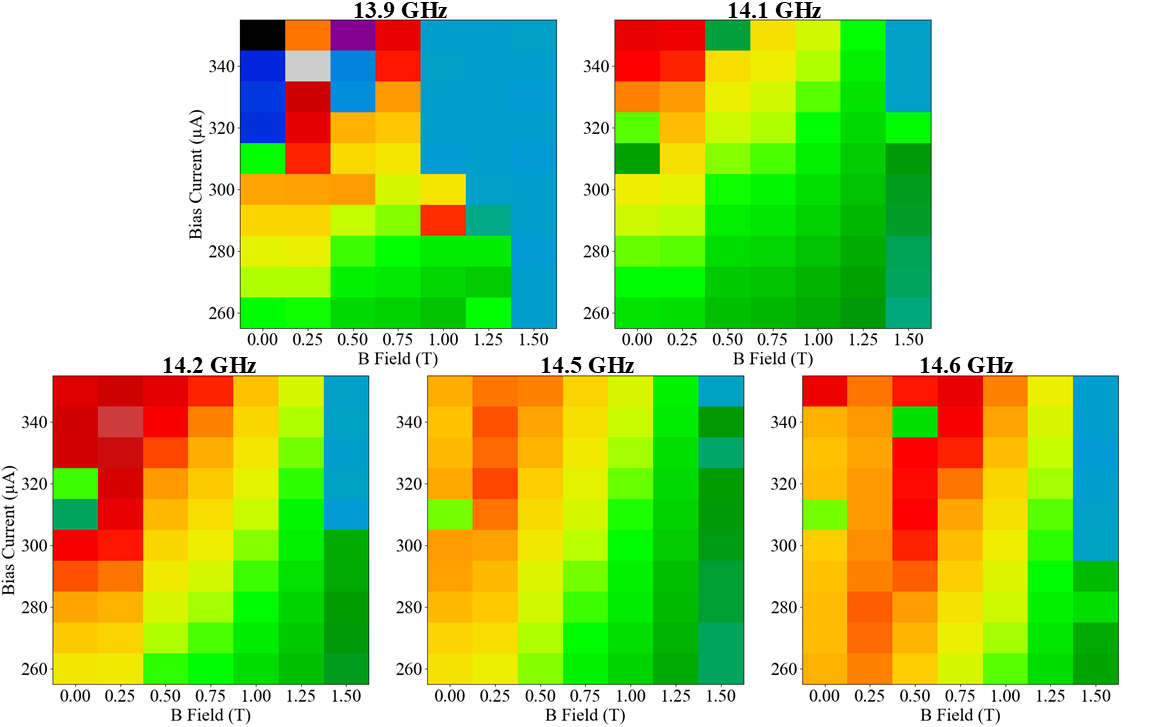}
    \includegraphics[width=0.3\textwidth]{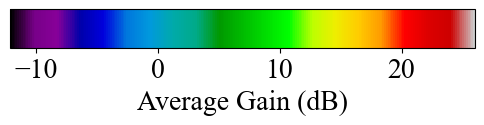}
    \end{center}
    \captionsetup{format=plain,justification=centering} 
    \caption{\justifying Colormaps representing average gain over different pump powers, represented for different bias currents as a function of magnetic field for KTWPA$_\parallel$. The pump powers are averaged with an exclusion that if two out of three pump powers tested produce positive gain but the third produces negative average gain, that the negative result is excluded as the device is no longer superconducting. Each panel represents a different pump frequency of KTWPA$_\parallel$. These colormaps highlight an observation that as bias current and pump frequency increases, the devices achieve higher gain for 0.25 T or 0.5 T than they do in no field. The field that can achieve highest gain in each panel increases as a function of the pump frequency, occurring under \SI{0.25}{\tesla} at 14.1~\si{GHz} and \SI{0.75}{\tesla} at 14.6~\si{GHz}.}
    \label{fig:migration}
\end{figure*}

To explore the role of magnetic field on device gain, we collapse the multi-dimensional space by fixing the device bias (current, pump power and frequency), and averaging over the gain bandwidth of 4~\si{GHz} to 10~\si{GHz} as a function of applied magnetic field, as shown in Fig.~\ref{fig:migration}, where we show only the KTWPA$_\parallel$ for brevity.  

At no applied B-field, the gain is maximized with a pump frequency of 14.2\,\si{GHz}. As the field is increased, the optimal bias point moves to higher pump frequencies. This behavior is consistent with a field-induced increase in the kinetic inductance of the amplifier. Consequently, the dispersion curve steepens and moves the signal-pump phase-matching condition to higher frequencies. In our operation, a 0.5\si{T} field requires a $\sim400\,\si{MHz}$ adjustment in tuning frequency. 
Fig.~\ref{fig:migration} illustrates that operation of an amplifier in a B-field requires a simple re-tuning of the pump-frequency bias point for optimal gain, and that gain is maintained even in the presence of a substantial magnetic field.

We have demonstrated in this letter the operation and gain performance of two KTWPAs of the design \cite{howe2025kineticinductancetravelingwave} over the $4~\si{GHz}$ to $10\,\si{GHz}$ frequency range in the presence of magnetic fields ranging in intensity over $0~\si{T}$ to $1.5\,\si{T}$. The KTWPA chips were oriented either normal or parallel to the introduced magnetic field, and were exercised over bias current and pump power \& frequency. 

By comparing our results to GL numerical solutions, we have determined that the loss of gain is unlikely to be due to suppression of the superconducting condensate under magnetic fields. While this occurs, the high $L_k$ material is robust to $B$-fields and gain is recovered by adjusting the pump to higher frequencies. 
Instead, a plausible model is that the magnetic field creates additional microwave dissipation consistent with GR vortex depinning dynamics. This counteracts the amplification along the transmission line, reducing the peak gain but roughly preserving the overall shape at a given operating point.  
However, the full frequency dependence is not fully described by a single depinning GR model. Nevertheless, these studies showed not only that KTWPAs can provide significant gain ($>$20~dB) in the presence of an introduced magnetic field, but can do so after having been exposed to multiple magnet ramps up to 2.5~Tesla without thermal quenching to expel trapped magnetic flux. 
The amplifiers exhibit peak gain at an applied 0.25~\si{mT} to 0.5~\si{mT} in-plane magnetic field and retain $\sim$10~dB of gain even under 1~\si{T} applied field.

A follow-up study is underway at the time of this writing to assess the effectiveness of KTWPAs of this design for application to axion haloscope searches, which require high intensity (multi-Tesla) magnetic fields to produce the telltale dark matter signal. The study will measure both the system gain and noise levels over multiple multi-Tesla magnet ramps, trapped flux quenches, in a mock axion haloscope search. Nonetheless, we are already encouraged by these results as they represent a significant preliminary step to making magnetic-field resilient quantum amplifiers available for basic science applications.



%
%

%

\section*{Acknowledgments}

This work was primarily supported by the U.S. Department of Energy, Office of Science, Office of High Energy Physics under Awards: DE-SC0026061 and C. Boutan's ECRP.
Additional support was provided by NIST Innovations in Measurement Science
program, the National Aeronautics and Space Administration (NASA) under Grant No. NNH18ZDA001N-APRA, and the Department of Energy (DOE) Accelerator and Detector Research Program under Grant No. 89243020SSC000058. 
This work is also supported by the DARTWARS project (EU H2020-MSCA Grant No. 101027746), and by the Italian NQSTI (PNRR MUR Grant No. PE0000023), and ICSC (PNRR MUR Grant No. CN00000013) quantum technologies programs.

D. Erdag and J. Van Vlack were supported in part by the U.S. Department of Energy, Office of Science, Office of Workforce Development for Teachers and Scientists (WDTS) under the Student Undergraduate Laboratory Internship program (SULI).

Certain commercial equipment, instruments, or materials are identified in this paper in order to specify the experimental procedure adequately. Such identification is not intended to imply recommendation or endorsement by the National Institute of Standards and Technology, nor is it intended to imply that the materials or equipment identified are necessarily the best available for the purpose. 

\bibliography{references}
\appendix
\section{Supplementary Information: Experimental Setup}
The tests described below were performed at Pacific Northwest National Laboratory (PNNL). Two KTWPAs were placed in a BlueFors LD400 dilution refrigerator (DR), mounted to the terminal end of an OFHC copper manifold as in Fig.~\ref{fig:schem}~B which itself was mounted to a gold-plated copper cold finger attached to the center of the DR mixing chamber (MXC) plate. The cold finger is sized and positioned such that KTWPAs are positioned near the center of the magnet bore. The magnet used is an American Magnetic Inc. Cryogen-free Compensated Vector Magnet with three independent NbTi coils: a z-axis solenoid about the vertical bore rated to 9~Tesla, and two smaller independent split coils to produce x-axis and y-axis fields of up to 1~Tesla each in the bore. The magnet is thermally sunk to the cryostat's 4~K stage but physically held below the MXC stage via a reinforced aluminum shield. Only the vertical field solenoid coil was controlled for these measurements.

The design of the copper manifold allows for one KTWPA to be mounted in a normal orientation to the direction of the bore's vertical magnetic field, which will be referred to here as
``KTWPA$_\perp$'', and a second KTWPA to be mounted in a parallel orientation, which will be referred to as
``KTWPA$_\parallel$''. 

Input and output RF lines are thermally sunk at each stage of the cryostat (so-called 50~K, 4~K, 1~K, 100~mK, and 10~mK, which were held at RT~K, 44~K, 3.8~K, 600~mK, $\sim$150~K, 10.05~mK with heating causing to MXC to reach up to 90~mK over the course of measurements) with input RF lines having 40~dB attenuation added to prevent saturation of the 10~mK stage with room temperature noise. Output lines are also thermally sunk and isolated to minimize high temperature noise from leaking into lower temperature stages, further amplified at the 4~K stage with a Low Noise Factory (LNF) HEMT (LNF-LNC0.3$\_$14B).

Further along the MXC stage on each KTWPA input is a K$\&$L 6L250 12~GHz low-pass filter used to suppress unwanted high frequency signals, followed next by a Marki BT-0018 Bias-T for combining the RF input signal and DC current bias, then a Krytar 120420 directional coupler, adding in the pump tone before reaching the KTWPA's effective input port. From the KTWPA's output port is another K$\&$L 6L250 low-pass filter, followed by another a Marki BT-0018 Bias-T for removing the DC bias, then a LNF-CICIC4$\_$12A dual junction cryogenic isolator to prevent noise from higher temperature stages from reaching the 10~mK devices. Both KTWPAs were pumped through a RF line that had only 10~dB of attenuation at the 4K stages to ensure adequate power on chip. Due to space constraints, this single line was shared by both KTWPAs and switched with a Raidall R577.432.000 mechanical switch at the mixing chamber stage. 

Room temperature equipment supporting input and readout includes a Yokogawa GS210 current source used to provide the DC bias, a SignalCore SC5511A tone generator to source the  pump signal, and a S5180B Copper Mountain Vector Network Analyzer (VNA) for testing RF transmission properties of the KTWPAs across the input and output lines.  

\begin{table*}[h!]
\centering
\begin{tabular}{|l||l|l|}
 \hline
 Dates & B-field (T) & Summary  \\
 \hline
 \hline
 Jan. 6 & 0 - 2.5 & KTWPA$_\perp$ / KTWPA$_\parallel$: Critical current measurements   \\
 \hline 
 Jan. 7 & 0 - 2.5 & KTWPA$_\parallel$: Parameter space investigation  \\
 \hline 
 Jan. 8 & 0 - 0.5 & KTWPA$_\perp$: Parameter space investigation   \\
 \hline 
 Jan. 9 & 0 - 1.5 & KTWPA$_\parallel$: Gain exploration dataset collection  \\
 \hline 
 Jan. 10 & 0 - 0.2 & KTWPA$_\perp$: Gain exploration dataset collection  \\
 \hline 
\end{tabular}
\\
\caption{Magnet ramp and measurement schedule in early 2025. }
\label{table:magramps}
\end{table*}

\end{document}